\documentclass[titlepage,pallis,twoside]{wpallis} %Exi dior8wseis meta tn dnmosieusn
\usepackage{fancyh}
\usepackage[l]{floatflt}
\usepackage{epsfig}
\usepackage{amssymb}
\usepackage{latexsym}
\usepackage{times}
\usepackage{slashed,cite}
\usepackage{upgreek}
\numberwithin{equation}{section}
\usepackage{rotating}
\usepackage{amsmath}
\usepackage{amsfonts}
\usepackage{amsbsy}
\usepackage{amscd}
\usepackage{bbm}

\newcommand{\crefs}[1]{Refs.~\cite{#1}}

\newcommand{\bal}{\begin{align}}
\newcommand{\eal}{\end{align}}
\newcommand{\beqs}{\begin{subequations}}
\newcommand{\eeqs}{\end{subequations}}
\newcommand{\eec}{\end{center}}
\newcommand{\bec}{\begin{center}}

\newcommand{\eem}{\end{matrix}}
\newcommand{\bem}{\begin{matrix}}
\newcommand{\eeq}{\end{equation}}
\newcommand{\beq}{\begin{equation}}
\newcommand{\ba}{\begin{array}}
\newcommand{\ea}{\end{array}}
\newcommand{\bea}{\begin{eqnarray}}
\newcommand{\eea}{\end{eqnarray}}
\newcommand{\baq}{\begin{eqnarray}}
\newcommand{\eaq}{\end{eqnarray}}

\newcommand\eqs[2]{Eqs.~(\ref{#1}) and (\ref{#2})}

\newcommand{\ftn}{\footnotesize}

\newcommand{\GeV}{{\mbox{\rm GeV}}}

\newcommand{\sFref}[2]{Fig.~\ref{#1}-{\ftn\sf ({#2})}}

\newcommand{\etal}{{\it et al.\/}}

\def\to{\rightarrow}

\def\llgm{\left\lgroup}
\def\rrgm{\right\rgroup}
\def\lf{\left(}
\def\rg{\right)}
\newcommand\vev[1]{\langle {#1} \rangle}

\newcommand{\Vhi}{\ensuremath{\widehat V_{\rm I}}}

\newcommand{\Hhi}{\ensuremath{\widehat H_{\rm I}}}
\newcommand{\Ohi}{\ensuremath{\Omega}}
\newcommand{\Omg}{\ensuremath{\Omega}}
\newcommand{\Khi}{\ensuremath{K}}

\newcommand{\Ns}{\ensuremath{{\what N_\star}}}
\newcommand{\ck}{\ensuremath{c_{T}}}

\newcommand{\mP}{\ensuremath{m_{\rm P}}}

\newcommand{\Qef}{\ensuremath{\Lambda_{\rm UV}}}

\def\openone{\leavevmode\hbox{\small1\kern-3.8pt\normalsize1}}
\newcommand{\dV}{\ensuremath{\Delta\widehat V_{\rm I}}}

\newcommand{\fp}{\ensuremath{f_\phi}}
\newcommand{\ft}{\ensuremath{f_T}}

\newcommand{\kx}{\ensuremath{k_S}}

\newcommand{\msn}{\ensuremath{\what m_{\rm \dph}}}

\newcommand{\ks}{\ensuremath{k_\star}}
\newcommand{\ns}{\ensuremath{n_{\rm s}}}
\newcommand{\na}{\ensuremath{{n_{11}}}}
\newcommand{\nb}{\ensuremath{n_{2}}}
\newcommand{\nc}{\ensuremath{n_{21}}}
\newcommand{\as}{\ensuremath{a_{\rm s}}}
\newcommand{\As}{\ensuremath{A_{\rm s}}}

\newcommand{\rcc}{\ensuremath{\mathcal{R}}}
\newcommand{\rce}{\ensuremath{\widehat{\mathcal{R}}}}
\newcommand{\Ve}{\ensuremath{\widehat{V}}}

\newcommand{\dphi}{\ensuremath{\what{\delta\phi}}}
\newcommand{\dph}{\ensuremath{\delta\phi}}

\newcommand{\what}{\ensuremath{\widehat}}
\newcommand{\wtilde}{\ensuremath{\widetilde}}

\def\aal{{\bar\alpha}}
\def\bbet{{\bar\beta}}
\def\al{{\alpha}}

\def\bt{{\beta}}

\def\W{{\widehat{W}}}
\def\K{{\widehat{K}}}

\def\n{\bar{n}}

\def\th{{\theta}}

\newcommand{\sg}{\ensuremath{\phi}}

\newcommand{\sgx}{\ensuremath{\phi_\star}}
\newcommand{\sgf}{\ensuremath{\phi_{\rm f}}}

\newcommand{\ld}{\ensuremath{\lambda}}

\newcommand{\Ld}{\ensuremath{\Lambda}}

\newcommand{\se}{\ensuremath{\widehat \phi}}
\newcommand{\sex}{\ensuremath{\widehat{\phi}_\star}}
\newcommand{\sef}{\ensuremath{\widehat{\phi}_{\rm f}}}
\newcommand{\geu}{\ensuremath{\widehat g}}
\newcommand{\eph}{\ensuremath{\widehat \epsilon}}
\newcommand{\ith}{\ensuremath{\widehat \eta}}

\newcommand\mtta[4]{\mbox{
$\llgm\bem #1 &#2 \cr #3& #4\eem\rrgm$}}

\def\Ka{K\"{a}hler potential}
\def\Km{K\"{a}hler manifold}
\def\Kaa{K\"{a}hler~}
\def\sub{subplanckian}

\def\str{Starobinsky}
\def\FHI{STI~}
\def\bcp{{\sc\small Bicep2}/{\it Keck Array}}
\newcommand{\plk}{{\it Planck}}

\newcommand{\diag}{\ensuremath{{\sf diag}}}

\begin{document}
\thispagestyle{empty}

%\preprint{UT-STPD-2/10}$SU(1,1)/U(1)\times SU(2)/U(1)$

\title[]{\Large\boldmath\bfseries\scshape Starobinsky-Type Inflation With Products \\ of K\"ahler Manifolds}

\author{\large\bfseries\scshape  C. Pallis \& N. Toumbas}
\address[] {\sl Department of Physics, University of Cyprus, \\ P.O. Box 20537,
CY-1678 Nicosia, CYPRUS \\  {\sl e-mail addresses: }{\ftn\tt
cpallis, nick@ucy.ac.cy}}

%Authors

\begin{abstract}{{\bfseries\scshape Abstract} \\
\par We present a novel realization of \str-type inflation within Supergravity using two
chiral superfields. The proposed superpotential is inspired by
induced-gravity models. The \Ka\ contains two logarithmic terms,
one for the inflaton $T$ and one for the matter-like field $S$,
parameterizing the $SU(1,1)/U(1)\times SU(2)/U(1)$ \Km. The two
factors have constant curvatures $-m/n$ and $2/\nb$, where $n$,
$m$ are the exponents of $T$ in the superpotential and \Ka\
respectively, and $0<\nb\leq6$. The inflationary observables
depend on the ratio $2n/m$ only. Essentially they coincide with
the observables of the original \str\ model. Moreover, the
inflaton mass is predicted to be $3\cdot10^{13}~\GeV$.}
\\ \\
{\ftn \sf Keywords: Cosmology of Theories Beyond the Standard
Model, Supergravity Models};\\
{\ftn \sf PACS codes:  98.80.Cq, 11.30.Qc, 12.60.Jv, 04.65.+e}
\\[0.2cm]
\publishedin{{\sl  J. Cosmol. Astropart. Phys.} {\bf 05}, no. 05,
015 (2016)}
\end{abstract} \maketitle

\setcounter{page}{1} \pagestyle{fancyplain}

%\addtolength{\headheight}{.5cm}$SU(1,1)/U(1)\times SU(2)/U(1)$

\rhead[\fancyplain{}{ \bf \thepage}]{\fancyplain{}{\sl
Starobinsky-Type Inflation With Products of K\"ahler Manifolds}}
\lhead[\fancyplain{}{\sl \leftmark}]{\fancyplain{}{\bf \thepage}}
\cfoot{}

\tableofcontents\vskip-1.3cm\noindent\rule\textwidth{.4pt}\\
%\vspace*{.3cm}

\section{Introduction}\label{intro}

The clarifications regarding the impact that the dust foreground
has on the observations of the B-type polarization of the CMBR,
offered by the recent joint analysis of the \bcp\ and \plk\ data
\cite{gws,plcp}, revitalizes the interest in the \str\ model
\cite{R2} of inflation. This model predicts a (scalar) spectral
index $\ns\simeq0.965$, which is in excellent agreement with
observations, and a tensor-to-scalar ratio $r\simeq0.0035$, which
is significantly lower than the current upper bound. Indeed, the
fitting of the data with the $\Lambda$CDM$+r$ model restricts
\cite{plcp} $\ns$ and $r$ in the following ranges
\beq \label{data}
\ns=0.968\pm0.0045\>\>\>\mbox{and}\>\>\>r\lesssim0.12 \eeq
at $95\%$ \emph{confidence level} ({\sf\ftn c.l.}), with negligible
$\ns$ running: $|\as|\ll0.01$.

On the other hand, \emph{Supergravity} ({\ftn\sf SUGRA})
extensions of the \emph{\str-type inflation} ({\sf\ftn STI}),
admit a plethora of incarnations \cite{ketov, matterlike, eno5,
eno7, linde, zavalos, R2r, tamvakis}. In most of them  two chiral
superfields, $T$ and $S$ are employed following the general
strategy introduced in \cref{referee} for the models of chaotic
inflation. One prominent idea \cite{eno5,eno7} is, though, to
parameterize with $S$ and $T$ the $SU(2,1)/(SU(2)\times U(1))$
\Km\ with constant curvature $-2/3$, as inspired by the no-scale
models \cite{noscale,lahanas}. In this context, a variety of
models have been proposed in which the inflaton can be identified
with either the matter-like field $S$ \cite{eno5, eno7,
matterlike} or the modulus-like field $T$ \cite{linde, eno7,
zavalos, R2r, tamvakis}. We shall focus on the latter case since
this implementation requires a simpler superpotential, and when
connected with a MSSM version, ensures a low enough re-heating
temperature, potentially consistent with the gravitino constraint
\cite{R2r,rehEllis,rehyoko}.

A key issue in such SUGRA realizations of \str\ inflation is the
stabilization of the field $S$ accompanying the inflaton. Indeed,
when the symmetry of the aforementioned \Km\ is respected, the
inflationary path turns out to be unstable against the
fluctuations of $S$. The instabilities can be lifted if we add to
the \Ka\ $K$ a sufficiently large quartic term $\kx|S|^4$, where
$\kx>0$ and $|\kx|\sim1$, as suggested in \cref{lee} for models of
non-minimal (chaotic) inflation \cite{linde1} and applied
extensively to this kind of models. This solution, however,
deforms slightly the \Km\ \cite{nick} and is complicated to
implement when more than two fields are present. In principle, all
allowed quartic terms have to be considered, rendering the
fluctuation analysis tedious -- see e.g. \cref{talk}.
Alternatively, we may utilize a nilpotent superfield $S$
\cite{nil}, or a matter field $S$ charged under a gauged
R-symmetry\cite{nick}.

We propose a new solution to the stability problem that is
compatible with a highly symmetric \Km. The \Ka\ involves a
logarithmic function of the inflaton field $T$ with an overall
negative prefactor, as required for establishing an asymptotic
inflationary plateau \cite{eno7, linde,zavalos, R2r}. If the
$|S|^2$ term is to appear in the argument of this logarithmic
function, its coefficient must be negative in order to avoid
negative kinetic terms. However such a negative coefficient leads
to tachyonic instabilities. Therefore, we propose to split $K$
into a sum of two logarithmic functions, one involving the
inflaton field $T$ and the other involving the field $S$, with
negative and positive prefactors $(-\na)$ and $\nb$, respectively.
The term $|S|^2$  can now appear in the argument of the second
logarithm with a positive coefficient. The prefactors $(-\na)$ and
$\nb$ are selected in order to establish STI, with the field $S$
acquiring a large enough, positive mass squared along the
inflationary trajectory. The resulting \Ka\ gives rise to the
product space $SU(1,1)/U(1)\times SU(2)/U(1)$.

We would like to comment on the possibility of realizing this type
of K\"ahler metrics in the context of string theory. The
non-compact coset factor, $SU(1,1)/U(1)$, appears in several
string induced no-scale models \cite{noscale, nick}.  There are
various classes of string inflationary models, namely D-brane
inflation in warped (and unwarped) superstring compactifications,
fluxbrane inflation, axion inflation, racetrack models, fibre
inflation and others -- see \cref{liam} for a thorough review and
references therein. In models with a D-brane, there are moduli
describing its position in the compactification manifold. Naively
one would think that the full moduli space is a product space. The
first factor, which is spanned by the brane position moduli, would
be isomorphic to the internal compact manifold, and the second
factor is a non-compact space spanned by the closed string moduli
(such as the modulus controlling the size of the internal space).
One could seek models in which the role of the inflaton is played
by a closed string modulus, or as in \cref{KKLMLT} a brane
position modulus. However, the stabilization of several closed
string moduli requires the presence of non-trivial fluxes. And
typically mixing arises between the brane position moduli with the
closed string K\"ahler moduli -- see \cref{KKLMLT,liam} for
discussions. As a result, the closed string moduli space is
fibered non-trivially over the space spanned by the brane position
moduli, as exemplified by the DeWolfe-Giddings K\"ahler potential
\cite{DwG, KKLMLT}. If the internal, compactification manifold
contains a spherical $SU(2)/U(1)$ factor, this must be supported
by suitable 2-form flux, which might affect the brane worldvolume
theory. Given this discussion, it may be difficult to realize a
situation in which the field configuration manifold is {\it
globally} isomorphic to the symmetric product space
$SU(1,1)/U(1)\times SU(2)/U(1)$ in the context of string
inflationary models. But at least {\it locally} in certain
regions, the moduli space could be approximated by a product space
of such form. This would require to turn on suitable fluxes in
order to stabilize some of the moduli in these regions. As argued
in \cref{KKLMLT,liam}, such a stabilization mechanism is likely to
steepen the inflaton potential, halting inflation. It is thus
challenging (and also interesting) to explicitly realize such a
model in the context of string theory.

%Such coset spaces can appear as factors of moduli spaces
%associated with superstring compactifications. The compact coset,
%for example, could arise as part of the open string moduli space,
%with the associated scalar fields parameterizing the positions of
%branes in the compactification manifold. It would be interesting
%to explicitly realize such a model in the context of string
%theory.

We implement our proposal within the framework \cite{old, gian,
rena, nIG} of \emph{induced-gravity} ({\sf\ftn IG}) models, which
are generalized to highlight the robustness of our approach. The
key-ingredient of our construction is the presence of the two
different exponents $n$ and $m$ of $T$  in the superpotential and
the \Ka. We show that imposing a simple asymptotic condition on
$n, m$ and $\na$, a \str-type inflationary potential gets
generated, exhibiting an attractor behavior that depends only on
the coefficient $\na$, which determines the curvature of the
$SU(1,1)/U(1)$ \Km. Moreover, this model of inflation preserves a
number of attractive features: {\sf\ftn (i)} The superpotential
and the \Ka\ may be fixed in the presence of an $R$-symmetry and a
discrete symmetry; {\sf\ftn  (ii)} the initial value of the
(non-canonically normalized) inflaton field can be \sub; {\sf\ftn
(iii)} the radiative corrections remain under control; and
{\sf\ftn (iv)} the perturbative unitarity is respected up to the
reduced Planck scale \cite{gian, riotto, R2r, nIG}.

The paper is organized as follows: In Sec.~\ref{fhi} we generalize
the formulation of \FHI within SUGRA IG models. In Sec.~\ref{fhi1}
we investigate totally symmetric \Ka s in order to find a viable
inflationary scenario, which is confronted with observations in
Sec.~\ref{fhi2}. Our conclusions are summarized in Sec.~\ref{con}.
Some mathematical notions related to the geometric structure of
the \Km s encountered in our set-up are exhibited in
Appendix~\ref{math}. Finally, Appendix~\ref{eff} provides an
analysis of the ultraviolet behavior of our models. Throughout,
charge conjugation is denoted by a star ($^*$), the symbol $,z$ as
subscript denotes derivation \emph{with respect to} ({\ftn\sf
w.r.t}) $z$ and we use units where the reduced Planck scale $\mP =
2.43\cdot 10^{18}~\GeV$ is equal to unity.

%\newpage

\section{Generalizing the Induced-Gravity Set-up in SUGRA}\label{fhi}

The realization of STI within IG models \cite{eno7, linde, R2r,
nIG, rena} requires the presence of two gauge singlet chiral
superfields, the inflaton $T$ and a ``stabilizer'' superfield $S$,
which we collectively denote by $z^\al$ ($z^1 =T$ and $z^2 =S$).
The relevant part of the  \emph{Einstein frame} ({\sf\ftn EF})
SUGRA action is given by \cite{linde1}
\beqs \beq\label{Saction1}  {\sf S}=\int d^4x \sqrt{-\what{
\mathfrak{g}}}\lf-\frac{1}{2}\rce +K_{\al\bbet}
\geu^{\mu\nu}\partial_\mu z^\al \partial_\nu z^{*\bbet}-\Ve\rg\,
\eeq
%
%where summation is taken over the scalar fields $z^\al$,
where the scalar field components of the superfields $z^\al$'s are
denoted by the same superfield symbol, $K_{\al\bbet}=K_{,z^\al
z^{*\bbet}}$ is the K\"ahler metric and $K^{\al\bbet}$ its inverse
($K^{\al\bbet}K_{\bbet\gamma}=\delta^\al_{\gamma}$). $\Ve$ is the
Einstein frame F--term SUGRA potential, given in terms of the \Ka\
and the superpotential $W$ by the following expression
\beq \Ve=e^{\Khi}\left(K^{\al\bbet}D_\al W D^*_\bbet W^*-3{\vert
W\vert^2}\right),\label{Vsugra} \eeq \eeqs
where $D_\al W=W_{,z^\al} +K_{,z^\al}W$. Next we perform a
conformal transformation \cite{linde1,lazarides} and define the
\emph{Jordan frame} ({\ftn\sf JF}) metric $g_{\mu\nu}$ via the
relation
\beqs\beq \label{weyl} \geu_{\mu\nu}=-\lf{\Omega}/{N}\rg
g_{\mu\nu},\eeq
where $\Omega$ is a frame function. In the JF, the action takes
the form
\beq {\sf S}=\int d^4x \sqrt{-\mathfrak{g}}\lf\frac{\Omega}{2N}
\rcc+\frac{3}{4N\Omega}\partial_\mu\Omega \partial^\mu\Omega
-\frac{1}{N}\Omega K_{\al{\bbet}}\partial_\mu z^\al \partial^\mu
z^{*\bbet}-V\rg\>\>\>\mbox{with}\>\>\>V=\frac{\Omg^2}{N^2}\Ve\,.\label{action2}\eeq\eeqs
Here $\mathfrak{g}$ stands for the determinant of $g_{\mu\nu}$;
$\rcc$ is the Ricci scalar curvature in JF, and $N$ is a
dimensionless positive parameter that quantifies the deviation
from the standard set-up \cite{linde1}. Let the frame function
$\Omega$ and $K$ be related by the equation
\beqs\beq-\Omega/N =e^{-K/N }\>\Rightarrow\>K=-N
\ln\lf-\Omega/N\rg\label{Omg1}.\eeq
Then using the on-shell expression \cite{linde1} for the purely
bosonic part of the auxiliary field
\beq {\cal A}_\mu =i\lf K_\al \partial_\mu z^\al-K_\aal
\partial_\mu z^{*\aal}\rg/6, \label{Acal1}\eeq
we arrive at the action
\beq {\sf S}=\int d^4x \sqrt{-\mathfrak{g}}\lf\frac{
\Omega}{2N}\rcc+\lf\Omega_{\al{\bbet}}+\frac{3-N}{N}
\frac{\Omega_{\al}\Omega_{\bbet}}{\Omega}\rg \partial_\mu z^\al
\partial^\mu z^{*\bbet}- \frac{27}{N^3}\Omega{\cal A}_\mu{\cal A}^\mu-V
\rg. \label{Sfinal}\eeq
In terms of $\Omega$, the auxiliary field ${\cal A}_\mu$ is given
by
\beq {\cal A}_\mu =-iN \lf \Omega_\al \partial_\mu
z^\al-\Omega_\aal
\partial_\mu z^{*\aal}\rg/6\Omega\,\label{Acal}\eeq\eeqs
where $\Omega_\al=\Omega_{,z^\al}$ and
$\Omega_\aal=\Omega_{,z^{*\aal}}$. This last form for the JF
action exemplifies the non-minimal coupling to gravity, as
$-\Omega/{N}$ multiplies the Ricci scalar $\rcc$. Conventional
Einstein gravity is recovered at the vacuum when
\beq -\vev{\Ohi}/N\simeq1. \label{ig}\eeq

Starting with the JF action in \Eref{Sfinal}, we seek to realize
STI, postulating the invariance of $\Ohi$ under the action of a
global $\mathbb{Z}_m$ discrete symmetry. When $S$ is stabilized at
the origin, we write
\beq -\Ohi/N=\Omega_{\rm H}(T)+\Omega^*_{\rm
H}(T^*)\>\>\>\mbox{with}\>\>\>\Omega_{\rm H}(T)= \ck
{T^m}+\sum_{k=2}^\infty\lambda_{k}{T^{km}}, \label{Omdef}\eeq
where $k$ is a positive integer. If the values of $T$ during
inflation are \sub\ and assuming relatively low $\lambda_{k}$'s,
the contributions of the higher powers of $T$ in the expression
above are very small, and so these can be dropped. As we will
verify later, this can be achieved when the coefficient $c_T$ is
large enough. Equivalently, we may rescale the inflaton setting $T
\to \tilde{T}={c_T}^{1/m}T$. Then the coefficients $\lambda_{k}$
of the higher powers in the expression of $\Ohi$ get suppressed by
factors of $c_T^{-k}$. Thus $\mathbb{Z}_m$ and the requirement
that the inflaton $T$ is \sub\ determine the form of $\Ohi$,
avoiding a severe tuning of the coefficients $\lambda_{k}$.
Confining ourselves to such situations (and stabilizing $S$ at the
origin), \Eref{Omg1} implies that the \Ka s take the form
\beq K_0=-N\ln\Big(
f(T)+f^*(T^*)\Big)\>\>\>\mbox{with}\>\>\>f(T)\simeq\ck T^m \,.
\label{K0} \eeq
\eqs{Omg1}{ig} require that $T$ and $S$ acquire the following
vacuum expectation values
\beq \vev{T}\simeq1/(2\ck)^{1/m},\>\>\>\mbox{and}\>\>\>
\vev{S}=0\,.\label{ig1} \eeq
These values can be obtained, if we choose the following
superpotential \cite{nIG,rena}:
\beq \label{Wn} W=\ld S\lf T^n-1/(2\ck)^{n/m}\rg\,, \eeq
since the corresponding F-term SUSY potential, $V_{\rm SUSY}$, is
found to be
\beq V_{\rm SUSY}= \ld^2\left|T^n- 1/(2\ck)^{n/m}\right|^2 +
\ld^2n^2\left|ST^{n-1}\right|^2 \label{VF}\eeq
and is minimized by the field configuration in \Eref{ig1}.
Similarly to \crefs{R2r,nIG}, we argue that when the exponent $n$
takes integer values with $n>1$, the form of $W$ is constrained if
we limit $T$ to \sub\ values, and if it respects two symmetries:
{\sf\ftn (i)} an $R$ symmetry under which $S$ and $T$ have charges
$1$ and $0$; {\sf\ftn (ii)} a discrete symmetry $\mathbb{Z}_n$
under which only $T$ is charged. For $n=m$, $\mathbb{Z}_m$ becomes
a symmetry of the theory and our scheme is essentially identical
to those analyzed in \crefs{rena, nIG}. Generalizing these
settings by allowing $n\neq m$, we find inflationary solutions for
a variety of combinations of the parameters $n,m$ and $N$ -- see
\Sref{res2} -- including the choice $N=3$ which appears in the
no-scale SUGRA models \cite{noscale,lahanas, eno5,eno7}. Note,
finally, that the selected $\Omega$ in \Eref{Omdef} does not
contribute in the term involving $\Omega_{TT^*}$ in \Eref{Sfinal}.
We expect that our finding are essentially unaltered even if we
include in the right-hand side of \Eref{Omdef} a term
$-(T-T^*)^2/2N$ \cite{rena} or $-|T|^2/N$ \cite{nIG} which yields
$\Omega_{TT^*}=1\ll\ck$. In those cases, however, the symmetry of
the \Km s, studied in \Sref{fhi1}, regarding the $T$ sector of the
models is violated.

The inflationary trajectory is determined by the constraints
\beq \label{inftr} S=T-T^*=0,\>\>\>\mbox{or}\>\>\>s=\bar
s=\th=0,\eeq
with the last equation arising when we parameterize $T$ and $S$ as
follows
\beq T=\:{\phi\,e^{i \th}}/{\sqrt{2}},\>\>\>\>\>\>S=\:(s +i\bar
s)/\sqrt{2}\,.\label{cannor} \eeq
Using the superpotential in \Eref{Wn}, we find via \Eref{Vsugra}
that, along the inflationary path, $\Ve$ takes the following form:
\beq \label{1Vhio}\Vhi=\Ve(\th=s=\bar s=0)=e^{K}K^{SS^*}\,
|W_{,S}|^2\,.\eeq
To identify the canonically normalized scalar fields, we cast
their kinetic terms in \Eref{Saction1} into the following diagonal
form
\beqs\beq \label{K} K_{\al\bbet}\dot z^\al \dot
z^{*\bbet}=\frac12\lf\dot{\se}^{2}+\dot{\what
\th}^{2}\rg+\frac12\lf\dot{\what s}^2 +\dot{\what{\overline
s}}^2\rg,\eeq
where the dot denotes derivation w.r.t the cosmic time and the
hatted fields are given by
\beq  \label{canp} {d\widehat \sg/ d\sg}=\sqrt{K_{TT^*}}=J,\>\>\>
\what{\th}= J\,{\th}/{\phi},\>\>\>(\what s,\what{\bar
s})=\sqrt{K_{SS^*}} {(s,\bar s)}.\eeq\eeqs
Note that the spinor components $\psi_T$ and $\psi_S$ of the $S$
and $T$ superfields must be normalized in a similar manner, i.e.,
$\what\psi_{S}=\sqrt{K_{SS^*}}\psi_{S}$ and
$\what\psi_{T}=\sqrt{K_{TT^*}}\psi_{T}$.

It is obvious from the considerations above, that the
stabilization of $S$ during and after inflation is of crucial
importance for the realization of our scenario. This issue is
addressed in the next section, where we specify the dependence of
the \Ka\ on $S$.

\section{\str-Type Inflation \& \Kaa Manifolds}\label{fhi1}

We focus on \Ka s parameterizing totally symmetric manifolds
consistent with the $R$ symmetry acting on $S$. In Sec.~\ref{21}
we review the models based on the $SU(2,1)/(SU(2)\times U(1))$
coset space. Then we analyze \Ka s parameterizing specific product
spaces: the $SU(1,1)/U(1)\times U(1)$ space in \Sref{111} and the
$SU(1,1)/U(1)\times SU(2)/U(1)$ space in \Sref{112}. Among these
cases, only the last one yields a satisfactory scenario.

\subsection{$SU(2,1)/(SU(2)\times U(1))$ \Kaa\ Manifold}
\label{21}

A typical \Ka\ employed for implementing \FHI in SUGRA is
\beq K_1=-\nc\ln\lf f(T)+f^*(T^*)-{|S|^2\over\nc}\rg,\label{K1}
\eeq
with $\nc>0$. The K\"ahler metric $K_{\al\bbet}$ takes the form
\beq \lf K_{\al\bbet}\rg=m\ck e^{2K_1/\nc}\mtta{m\nc\ck
|T|^{2(m-1)}}{-ST^{m-1}}{-S^*T^{*(m-1)}}{\lf
T^m+T^{*m}\rg/m}.\label{Kab1} \eeq
Using this expression, the superpotential of \Eref{Wn} and
\Eref{Vsugra}, we obtain:
\bea\Ve&=&\frac{\ld^2e^{K_1}}{(2\ck)^{2n/m}m^2\nc^2\ck|T|^{2m}}
\Bigg(m^2\ck^2\nc^2|T|^{2m}\lf T^{m}+T^{*m}\rg|\ft|^2 \nonumber
\\&-&(2\ck)^{n/m}|S|^4\lf(2\ck)^{n/m}n|T|^{2n} \lf
T^{m}+T^{*m}\rg+m(\nc-1) \lf T^{m}T^{*n}\ft+T^{*m}T^{n}\ft^*\rg
\rg\nonumber  \\&+&\nc\ck|S|^2\lf(2\ck)^{2n/m}n^2|T|^{2n} \lf
T^{m}+T^{*m}\rg^2+ m^2(\nc^2-3\nc-1)|T|^{2m}|\ft|^2\right.
\nonumber  \\&+&\left. (2\ck)^{n/m}nm(\nc-1)\lf T^{m}+T^{*m}\rg\lf
T^{m}T^{*n}\ft+T^{*m}T^{n}\ft^*\rg\rg\Bigg)\,, \label{Ve1}\eea
where $\ft=1-\lf2\ck\rg^{n/m}T^n$. Along the inflationary track in
\Eref{inftr}, $K_{\al\bbet}$ becomes diagonal
%\succapprox
\beq \lf K_{\al\bbet}\rg= \diag\lf {\nc
m^2\over2\sg^2},{2^{n/2}\over2\ck\sg^n}\rg, \label{K1a}\eeq
while \Eref{Ve1} reduces to \Eref{1Vhio}, given explicitly
by
\beq
\Vhi=\frac{2^{-n+(m-2)(\nc-1)/2}\ld^2\fp^2}{\ck^{2n/m+\nc-1}\sg^{m(\nc-1)}}\>\>\>\>
\mbox{with}\>\>\>\>\fp=2^{-n/m+n/2}\ft\,.\label{Vhi1} \eeq
The function $\ft$ becomes a function of $\sg$ along the
inflationary trajectory -- see \Eref{cannor}. When $\ck\gg1$ and
$\sg<1$, or $\ck=1$ and $\sg\gg1$, $\Vhi$ develops a plateau with
almost constant potential energy density, if the exponents are
related as follows
\beq \label{con1}
2n=m(\nc-1)\>\>\>\Rightarrow\>\>\>m=2n/(\nc-1)\,.\eeq
For $m=n$, \Eref{con1} yields $\nc=3$, which is the standard
choice -- cf. \cref{nIG}. Moreover, if we set $m=n=\ck=1$ and
$\nc=3$, $W$ and $K_1$ in \eqs{Wn}{K1} yield the model of
\cref{cec}, which is widely employed in the literature \cite{eno7,
zavalos, linde} for implementing STI within SUGRA. As we verified
numerically, the data on $\ns$ -- see \Eref{data} -- permit only
tiny (of order $0.001$) deviations from \Eref{con1}, in accordance
with the findings of \cref{tamvakis}. More pronounced (of order
$0.01$) deviations have been found to be allowed in \cref{np1},
where a higher order mixing term $|S|^2|T|^2$ is considered. In a
such case, a sizable increase of $r$ can be achieved, but the
symmetry of the \Km\ is violated. Since integers are considered as
the most natural choices for $\nc, n$ and $m$, we adopt throughout
conditions like the above one as empirical criteria for obtaining
observationally acceptable STI.

Eliminating $m$ via \Eref{con1}, $\Vhi$ and $\fp$ in \Eref{Vhi1}
are written as -- cf. \cref{nIG}:
\beq \label{Vhi1o}\Vhi=\frac{2^{1-\nc} \ld^2\fp^2}{\ck^{2(\nc-1)}
\sg^{2n}}\>\>\>\>\mbox{with}\>\>\>\>\fp=2^{(1+n-\nc)/2}
-\ck^{(\nc-1)/2}\sg^n\,.\eeq
Integrating the first equation in \Eref{canp}, we can find the
EF canonically normalized field $\se$ as a function of $\sg$.
We can then express $\Vhi$ in terms of $\se$ obtaining
\beq \label{Vhie}
\Vhi(\se)=\frac{2^{1-\nc}\ld^2}{\ck^{\nc-1}}\lf1-
e^{-\frac{1-\nc}{\sqrt{2\nc}}\,\se}\:\rg^2
\>\>\>\mbox{with}\>\>\>\se=\frac{\sqrt{2\nc}}{\nc-1}n\ln\lf(2\ck)^{(\nc-1)/2n}\frac{\sg}{\sqrt{2}}\rg,\eeq
where the integration constant is evaluated so that
$\Vhi(\se=0)=0$. When $n=\ck=1$ and $\nc=3$, $\Vhi$ coincides with
the potential extensively used in the realizations of STI. It is
well-known, however, that the inflationary trajectory is unstable
against the fluctuations of $S$ \cite{linde1,linde}. In Table~1,
we display the mass-squared spectrum along the trajectory in
\Eref{inftr} for the various choices of $K$. When $K=K_1$, we find
$\what m^2_{s}<0$, since the result is dominated by the negative
term $-\ck^{\nc-1}\sg^{2n}$. This occurs even when $\nc=1$. Note
that there are no instabilities along the $\theta$ direction,
since $\what m^2_{\theta}/\Hhi^2>1$, where $\Hhi^2=\Vhi/3$ is the
Hubble parameter squared, and $\Vhi$ is estimated by \Eref{Vhi1o}.
In Table~1, we also list the masses $\what m^2_{\psi^\pm}$ of the
fermion mass-eigenstates $\what \psi_\pm=(\what{\psi}_{T}\pm
\what{\psi}_{S})/\sqrt{2}$ given in terms of the canonically
normalized spinors defined in \Sref{fhi}.

\subsection{$SU(1,1)/U(1)\times U(1)$ \Kaa\ Manifold}
\label{111}

As shown in \cref{lazarides}, in a similar set-up, the situation
regarding the stability along the $S$ direction can be improved if
we choose a different \Ka:
\beq K_2=-\na\ln\Big(f(T)+f^*(T^*)\Big)+|S|^2,\label{K2}\eeq
where $\na>0$. This \Ka\ parameterizes  \cite{tamvakis} the
$SU(1,1)/U(1)\times U(1)$ manifold. The $S$ field has a positive
mass squared $\what m^2_s$, but this turns out to be less than
$\Hhi^2$ -- see Table~1.

\setcounter{table}{1}
\renewcommand{\arraystretch}{1.6}%\vspace*{-5.0cm}
\begin{sidewaystable}
\vspace*{15.0cm}\begin{center}
\begin{tabular}{|c|c|c|c|c|c|}\hline
{\sc Fields}&{\sc Eigen-}& \multicolumn{4}{c|}{\sc Masses
Squared}\\\cline{3-6}
&{\sc states}&& \hspace*{0.cm}{$K=K_1$}&{$K=K_2$} &{$K=K_{3}$} \\
\hline\hline
1 real scalar&$\widehat\theta$&$\widehat m_{\theta}^2/\Hhi^2$&
$6(\nc-1)\lf2^{1+n}+2^{\nc}\ck^{\nc-1}\sg^{2n}+\right.$
&\multicolumn{2}{c|}{$\lf6/\fp^2\rg\lf2^{n-\na}+\ck^\na\sg^{2n}\right.$}\\
&&&$\left.2^{\frac12(n+\nc-1)}
(\nc-5)\ck^{\frac12(\nc-1)}\sg^n\rg/2^{\nc}\nc\fp^2$&\multicolumn{2}{c|}{$+\left.2^{\frac12(n-\na)-1}
(\na-4)\ck^{\na/2}\sg^n\rg$}\\\cline{4-6}
1 complex &$\widehat s, \widehat{\bar{s}}$ & $\widehat m_
s^2/\Hhi^2$&{$\lf6/\nc\fp^2\rg\lf2^{\frac12(3-\nc+n)}
\ck^{(\nc-1)/2}\sg^n\right.$}&$3\cdot2^{n-\na}\na/\fp^2$&
$3\lf2/\nb+2^{n-\na}\na/\fp^2\rg$\\
scalar&&&{$\left.-\ck^{\nc-1}\sg^{2n} +2^{n-\nc}\lf\nc(\nc-2)-1\rg
\rg$}&&\\\hline
$2$ Weyl spinors & $\what \psi_\pm $ & $\what m^2_{ \psi\pm}$
&$2^{n-2(\nc-1)}(\nc-1)^2\ld^2/\nc\ck^{2(\nc-1)}\sg^{2n}$
&\multicolumn{2}{c|}{\hspace*{0.5cm}$2^{n-2\na}\na\ld^2/\ck^{2\na}\sg^{2n}$}
\\\hline
\end{tabular}\\[0.5cm]%\captionsetup
{\slshape\bfseries \small Table~1:} {\sl\small Mass-squared
spectrum for $K=K_1, K_2$ and $K_3$ along the direction in
\Eref{inftr}.}
\end{center}%\hspace*{5cm}\captionof{table}{}
\end{sidewaystable}

In this model the \Kaa metric is diagonal for any value of $T$ and
$S$, i.e.,
%\succapprox
\beq \label{Kab2} \lf K_{\al\bbet}\rg= \diag\lf {\na
m^2|T|^{2(m-1)}\over\lf T^m+T^{*m}\rg^2},1\rg\,.\eeq
Inserting the above result and $W$ in \Eref{Wn} into
\Eref{Vsugra}, we arrive at
\bea \nonumber \Ve&=&\frac{\ld^2e^{|S|^2}}{ \lf2\ck\rg^{2n/m}
\ck^\na \lf T^{m}+T^{*m}\rg^\na}
\Bigg(\lf1+|S|^2\rg^2|\ft|^2-3|S|^2|\ft|^2\\
&+& \frac1{m^2\na}|S|^2|T|^{-2m} \left|m\na
T^m\ft+(2\ck)^{n/m}nT^n\lf
T^{m}+T^{*m}\rg\right|^2\Bigg)\,.\label{Ve2}\eea

Along the inflationary path, \eqs{Kab2}{Ve2} simplify as follows
%\succapprox
\beq \label{Vhi2} \mbox{\small\sf (a)}\>\>\lf K_{\al\bbet}\rg=
\diag\lf {\na\,
m^2\over2\sg^2},1\rg\>\>\>\mbox{and}\>\>\>\mbox{\small\sf
(b)}\>\>\Vhi=\frac{2^{-n+(m-2)\na/2}\ld^2\fp^2}{\ck^{2n/m+\na}\sg^{m\na}}\, ,\eeq
where $\fp$ coincides with the function defined in \Eref{Vhi1}, independently of $\na$.
The asymptotic condition which ensures \FHI is now expressed as --
cf.~\Eref{con1}:
\beq \label{con2} m\na=2n\>\>\>\Rightarrow\>\>\>m=2n/\na\,.\eeq
As shown in Appendix~\ref{math}, this condition gives the ratio of
the exponents $m$ and $n$ in terms of minus the curvature of the
$SU(1,1)/U(1)$ \Km\ in Planck units. For $n=m$, we end up with the
IG models considered in \cref{nIG} and \Eref{con2} yields $\na=2$.
Setting $\na=2(1+\bar n_1)$, we find that consistency with
\Eref{data}, regarding $\ns$, restricts $\bar n_1$ in a very
narrow region $-1/200\lesssim\bar n_1\lesssim1/250$. Since this
result indicates significant tuning, we do not pursue this
possibility.

In terms of $n$ and $\na$, $\Vhi$ in \Eref{Vhi2} takes the form
\beq\Vhi=\frac{\ld^2\fp^2}{2^{\na}\ck^{2\na}
\sg^{2n}}\>\>\>\mbox{with}\>\>\>\fp=2^{(n-\na)/2}
-\ck^{\na/2}\sg^n\cdot\label{Vhi2o}\eeq
As before we express
$\sg$ and $\Vhi$ in terms of the canonically normalized field $\se$:
\beq \Vhi(\se)=\lf2\ck\rg^{-\na}\ld^2\lf1-
e^{-\sqrt{{\na/2}}\,\se}\:\rg^2\>\>\>\mbox{with}\>\>\>\se=\sqrt{\frac{2}{\na}}
n\ln\lf(2\ck)^{\na/2n}\frac{\sg}{\sqrt{2}}\rg\,,\label{Vhiee}\eeq
where the integration constant satisfies the same condition as in
\Eref{Vhie}. The resulting expressions share similar qualitative
features with those expressions.

The relevant mass spectrum for the choice $K=K_2$ is shown in
Table~1. Although $\what m^2_{\chi^\al}>0$ for $\chi^\al=\th$ and
$s$, we observe that $\what m^2_{s}/\Hhi^2<1$ since $\fp^2\gg1$
for $\ck\gg1$ and $\sg<1$ (or $\sg\gg1$ and $\ck<1$). Here we take
$\Hhi^2=\Vhi/3$ with $\Vhi$ given by \Eref{Vhi2o}. This result
arises due to the fact that only the term in the second line of
\Eref{Ve2} contributes to $\what m^2_{s}$. Since there is no
observational hint \cite{plcp} for large non-Gaussianity in the
cosmic microwave background, we prefer to impose that $\what
m^2_s\gg\Hhi^2$ during the last $50-60$ e-foldings of inflation.
This condition guarantees that the observed curvature perturbation
is generated only by $\sg$, as assumed in \Eref{Prob} below.
Nonetheless, two-field inflationary models which interpolate
between the \str\ and the quadratic model have been analyzed in
\cref{2field}.

\subsection{$SU(1,1)/U(1)\times SU(2)/U(1)$ \Kaa\ Manifold}
\label{112}

To obtain a large mass for the fluctuations of $S$, we replace the
second factor of the product manifold of \Sref{111} with a compact
coset space. Thus, we consider the following \Ka\
\beq K_3=-\na\ln\Big(
f(T)+f^*(T^*)\Big)+\nb\ln\lf1+\frac{|S|^2}{\nb}\rg, \label{K3}\eeq
where $\nb>0$. \Eref{K3} together with \eqs{Wn}{Vsugra} imply that
along the inflationary direction in \Eref{inftr}, $K_{\al\bbet}$
and $\Vhi$ are given by the expressions in \Eref{Vhi2} and
$\Vhi(\se)$ by \Eref{Vhiee}. Therefore, the inflationary plateau
for \FHI is obtained by enforcing \Eref{con2}. Contrary to the
model of \Sref{111}, though, the fluctuations of $S$ turn out to
be adequately heavy, as shown in Table~1 for the choice $K=K_3$
and $0<\nb\leq6$.

Indeed, $K_{\al\bbet}$ now differs from that in \Eref{Kab2} w.r.t
its second entry, i.e.,
%\succapprox
\beq \label{Kab3} \lf K_{\al\bbet}\rg= \diag\lf {\na
m^2|T|^{2(m-1)}\over\lf
T^m+T^{*m}\rg^2},\lf1+\frac{|S|^2}{\nb}\rg^{-2}\rg\,.\eeq
Substituting $K_{\al\bbet}$ and $W$ from \Eref{Wn} into
\Eref{Vsugra}, we end up with
\bea \nonumber \Ve&=&\frac{\ld^2\lf1 + {|S|^2}/{\nb}\rg^{\nb}}{
\lf2\ck\rg^{2n/m} \ck^\na \lf T^{m}+T^{*m}\rg^\na}
\Bigg(\lf1+\lf1+\frac1\nb\rg|S|^2\rg^2|\ft|^2-3|S|^2|\ft|^2\\
&+& \frac1{m^2\na}|S|^2|T|^{-2m} \left|m\na
T^m\ft+(2\ck)^{n/m}nT^n\lf T^{m}+T^{*m}\rg\right|^2\Bigg)\,.
\label{Ve3}\eea
Comparing this last expression with that in \Eref{Ve2}, we see
that the first term in the parenthesis is enhanced by a factor
$(1+1/\nb)$. This is the origin of the additional $6/\nb$ term in
the expression of $\what m^2_s$ -- compare in Table~1 the mass
expressions for the choices $K=K_3$ and $K=K_2$. This extra term
dominates when $|\nb|\leq6$, yielding $\what m^2_s>\Hhi^2$ (for
$\nb>0$). On the contrary, for $\nb<0$ -- when the corresponding
\Km\ is $(SU(1,1)/U(1))^2$ -- taking values in the range
$-6<\nb<0$, the instability occurring for the $K=K_1$ choice
reappears. For $\nb<-6$, the mass squared may be positive but we
obtain $\what m^2_s<\Hhi^2$, as in the $K=K_2$ case. Note that the
bounds on $\nb>0$ constrain the curvature of the $SU(2)/U(1)$ \Km\
-- see Appendix~\ref{math}. Note also that, in contrast to
\Eref{Kab1}, the denominator of $K_{TT^*}$ in \Eref{Kab3} does not
depend on $S$. As a consequence, no geometric destabilization
\cite{renaux} can be activated in our model, differently to the
conventional case of STI realized by the $K=K_1$ choice.

\section{Inflation Analysis} \label{fhi2}

It is well known \cite{rena, nIG} that \FHI based on $\Vhi$ of
\Eref{Vhie}, with $n=m$ and $\nc=3$, exhibits an attractor
behavior in that the inflationary observables and the inflaton
mass at the vacuum are independent of $n$. It would be interesting
to investigate if and how this nice feature gets translated in the
extended versions of \FHI based on $\Vhi$ of \Eref{Vhiee}. In this
section we examine this issue. We test our models against
observations, first analytically in \Sref{res1} and then
numerically in \Sref{res2}.

\subsection{Analytic Results}\label{res1}

\paragraph{4.1.1 Duration of STI.} The number of e-foldings, $\Ns$, that
the pivot scale $\ks=0.05/{\rm Mpc}$ undergoes during inflation has
to be large enough to solve the horizon and flatness problems of the
standard Big Bag cosmology, i.e.,
\begin{equation} \label{Nhi} \Ns=\int_{\sef}^{\sex}\, d\se\:
\frac{\Vhi}{\Ve_{{\rm I},\se}}= \int_{\sgf}^{\sgx}\, d\sg\:
J^2\frac{\Ve_{\rm I}}{\Ve_{{\rm I},\sg}}\simeq(50-60).
\end{equation}
The precise numerical value depends on the height of the
inflationary plateau, the re-heating process and the cosmological
evolution following the inflationary era \cite{plcp}. Here
$\sgx~[\sex]$ is the value of $\sg~[\se]$ when $\ks$ crosses the
inflationary horizon. The other integration limit, $\sgf~[\sef]$,
is set by the value of $\sg~[\se]$ at the end of inflation. In the
slow-roll approximation, this is determined by the condition:
\beqs\beq \label{sr12} {\ftn\sf max}\{\what\epsilon(\sgf),\
|\what\eta(\sgf)|\}=1,\eeq where the slow-roll parameters, for
$\Vhi$ given in \Eref{Vhi2o}, are given by -- cf. \cref{nIG}:
\beq \label{sr1}\what \epsilon= {1\over2}\left(\frac{\Ve_{{\rm
I},\se}}{\Ve_{\rm
I}}\right)^2=\frac{2^{n-\na}\na}{\fp^2}\>\>\>\mbox{and}\>\>\>
\what\eta=\frac{\Ve_{{\rm I},\se\se}}{\Ve_{\rm
I}}={\frac{\na}{\fp^2}\lf2^{1+n-\na}-2^{\frac12(n-\na)}\ck^{\frac12\na}\sg^n\rg}\,.
\eeq\eeqs
Therefore, the end of inflation is triggered by the
violation of \Eref{sr12} at a value of $\sg$ given by the condition
\beq \label{sr12a} \sgf={\ftn\sf
max}\left\{\sqrt{2}\lf1+\sqrt{2}\over2\ck\rg^{1/n} ,\
2^{(n-\na)/2n}\lf2-\na+\sqrt{\na(\na+4)}\over2\ck^{\na/2}\rg^{1/n}\right\}\,\cdot\eeq

The integral in \Eref{Nhi} yields
\beqs\beq \label{N*}
\Ns={2^{(\na-n)/2}\over\na}\,\ck^{\na/2}\lf{\sgx^n-\sgf^n}\rg-\frac{n}{\na}\ln\frac{\sgx}{\sgf}\,\cdot\eeq
Ignoring the logarithmic term and taking into account that
$\sgf\ll\sgx$, we obtain a relation between $\sgx$ and $\Ns$:
\beq \label{sgx}
\sgx\simeq2^{(n-\na)/2n}\ck^{-\na/2n}\lf\na\Ns\rg^{1/n}\,.\eeq\eeqs
When $\ck=1$ the requirement of \Eref{Nhi} can be fulfilled only
for $\sgx\gg1$ -- see e.g. \crefs{linde, eno7, zavalos}. On the
contrary, letting $\ck$ vary, inflation can take place with
subplanckian $\sg$'s, since
\beq \label{ckmin}
\sgx\leq1\>\>\>\Rightarrow\>\>\>\ck\geq2^{n/\na-1} \lf \na
\Ns\rg^{2/\na}\,.\eeq
Therefore, we need relatively large values for $\ck$, which increase with $n$
and $1/\na$. As shown in Appendix~\ref{eff}, this feature of the
models does not cause any problem with perturbative unitarity,
since $\se$ in \Eref{Vhiee} does not coincide with $\sg$ at the
vacuum of the theory -- contrary to conventional non-minimal
chaotic inflation \cite{R2r, riotto, gian, nIG}.

\paragraph{4.1.2 Normalization of the power spectrum.}
The amplitude $\As$ of the power spectrum of the curvature
perturbation generated by $\phi$ at the pivot scale $k_\star$ is
to be confronted with the data~\cite{plcp}:
\beqs\begin{equation}  \label{Prob} A^{1/2}_{s}=\:
\frac{1}{2\sqrt{3}\, \pi^3} \; \frac{\Ve_{\rm
I}(\sex)^{3/2}}{|\Ve_{{\rm I},\se}(\sex)|}
=\frac{\ld(1-\na\Ns)^2}{2^{(3+\na)/2}\sqrt{3}\pi\ck^{\na/2}\na^{3/2}\Ns}
\simeq4.627\cdot 10^{-5}. \eeq
Since the scalars listed in Table~1 for the choice $K=K_3$, with
$0<\nb\leq6$, are massive enough during inflation, the curvature
perturbations generated by $\sg$ are solely responsible for
generating $\As$. Substituting \eqs{sr1}{sgx} into the above
relation, we obtain
\beq\ld\simeq2^{(3+\na)/2}\sqrt{3\As/\na}\pi\ck^{\na/2}/\Ns\>\>\Rightarrow\>\>
\ck\simeq\lf5.965\cdot10^9\ld^2\na\rg^{1/\na}/2\,,\label{lan}
\eeq\eeqs
for $\Ns\simeq55$. Therefore, enforcing \Eref{Prob}, we obtain a
constraint on $\ld/\ck^{\na/2}$ which, by virtue of \Eref{con2},
depends exclusively on $\na$. Note, however, that $\ck$ inherits
though \Eref{sgx} an $n$ depedence which is also propagated to
$\ld$ via \Eref{lan}.

\paragraph{4.1.3 Inflationary Observables.} The inflationary observables
can be estimated through the relations -- cf.~\cref{nIG}:
\beqs\baq \label{ns} \ns&=&\: 1-6\what\epsilon_\star\ +\
2\ith_\star=\frac{1+n^2_{11}(\Ns-2)\Ns-2\na(1+\Ns)}{(1-\na\Ns)^2}
\simeq1-{2\over\Ns}-{6\over\na\Ns^2}, \>\>\> \\
\label{as} \as
&=&\:{2\over3}\left(4\ith_\star^2-(\ns-1)^2\right)-2\what\xi_\star=-\frac{2n^3_{11}\Ns(\na\Ns+3)}{(1-\na\Ns)^4}
\simeq-{2\over\Ns^2}-{14\over\na\Ns^3},\>\>\> \\
\label{resr} r
&=&16\eph_\star=\frac{16\na}{(1-\na\Ns)^2}\simeq{16\over\na\Ns^2}+{32\over
n^2_{11}\Ns^3}, \eaq\eeqs
where $\what\xi={\Ve_{{\rm I},\se} \Ve_{{\rm
I},\se\se\se}/\Vhi^2}$, and the variables with subscript $\star$
are evaluated at $\phi=\sgx$. We observe that the analytic
expressions for $\ns,\as$ and $r$ depend exclusively on $\na$, and
therefore, they deviate from those obtained in \cref{nIG} for the
choice $K=K_1$ and $\nc=3$. However, their numerical values --
shown in \Tref{tab2} for $\Ns=55$ and various combinations of $n$
and $\na$ -- are essentially the same with those findings. Indeed,
the leading terms in the expansions in \eqs{ns}{as} are identical
with the corresponding ones in \cref{nIG}. Only $r$ turns out to
be more sensitive to the change from $\nc$ to $\na$. In any case
its value remains below $0.005$ for reasonable values of $\na$.

\paragraph{4.1.4 Mass of the inflaton.} The EF, canonically
normalized, inflaton
\beq\dphi=\vev{J}\dph\>\>\>\mbox{with}\>\>\>\vev{J}=\sqrt{\frac{2}{\na}}\frac{n}{\vev{\sg}}=
\frac{n}{\sqrt{\na}}\lf2\ck\rg^{\na/2n}\>\>\>\mbox{and}\>\>\>
\dph=\phi-\vev{\phi} \label{dphi} \eeq
acquires a mass, at the SUSY vacuum -- see \Eref{ig1} -- given by
\beq \label{msn} \msn=\left\langle\Ve_{\rm
I,\se\se}\right\rangle^{1/2}= \left\langle \Ve_{\rm
I,\sg\sg}/J^2\right\rangle^{1/2}
=\frac{\ld\sqrt{\na}}{\lf2\ck\rg^{\na/2}}\simeq\frac{2\sqrt{6\As}\pi}{\Ns}\,\cdot\eeq
Note that no SUSY breaking vacua, as those analyzed in
\cref{farakos}, are present in our set-up. It is remarkable that
$\msn$ is essentially independent of $n$ and $\na$ thanks to the
relation between $\ld$ and $\ck$ in \Eref{lan}. It is also
interesting that even if we had followed the same analysis for
$K=K_1$ in \Eref{K1} we would have found essentially the same mass
of the inflaton. In particular in that case we would have obtained
\beq \label{msn1}
\msn=\frac{\ld\lf\nc-1\rg}{\lf2\ck\rg^{\frac12(\nc-1)}\sqrt{\nc}}
=\frac{2 \sqrt{6\As}\pi(\nc-1)^4 \Ns}{(\nc - (\nc-1)^2
\Ns)^2}\simeq\frac{2\sqrt{6\As}\pi}{\Ns}\,\cdot\eeq
Therefore, our models are practically indistinguishable from other
versions of \FHI as regards $\msn$. In other words, the condition
in \Eref{con2} generates for every $\na$ a novel --
cf.~\crefs{rena,nIG} -- class of attractors in the space of the
\str-like inflationary models within SUGRA.

\subsection{Numerical Results}\label{res2}

The analytic results presented above can be verified numerically.
Let us recall that the inflationary scenario depends on the following
parameters -- see \eqs{Wn}{K3}:
$$m,\>n,\>\na,\>\nb,\>\ck,\>\mbox{and}\>\ld.$$
The first three are constrained by \Eref{con2}, whereas the fourth
does not affect the inflationary outputs, provided that $\what
m_{s}^2>\Hhi^2$ for every allowed $n, \na$ and $\ck$. This is
satisfied when $0<\nb\leq6$, as explained in \Sref{112}. The
remaining parameters together with $\sgx$ can be determined by
imposing the observational constraints in Eqs.~(\ref{Nhi}), for
$\Ns=55$, and (\ref{Prob}). Note that in our code we find $\sgx$
numerically without the simplifying assumptions used for deriving
\Eref{sgx}. Moreover, \Eref{ckmin} bounds $\ck$ from below,
whereas \Eref{Prob} provides a relation between $\ck$ and $\ld$,
as derived in \Eref{lan}. Finally, we employ the definitions of
$\ns, \as$ and $r$ in Eqs.~(\ref{ns}) -- (\ref{resr}) to extract
the predictions of the models and \Eref{msn} to find the inflaton
mass.

In our numerical computation, we also take into account the
one-loop radiative corrections, $\dV$, to $\Vhi$ obtained from the
derived mass spectrum -- see Table~1 -- and the well-known
Coleman-Weinberg formula. It can be verified that our results are
insensitive to $\dV$, provided that the renormalization group mass
scale $\Lambda$ is determined by requiring $\dV(\sgx)=0$ or
$\dV(\sgf)=0$. A possible dependence of the results on the choice
of $\Lambda$ is totally avoided thanks to the smallness of $\dV$
for any $\nb$ with $0<\nb\leq6$, giving rise to
$\Ld\simeq(1-1.8)\cdot10^{-5}$ -- cf. \cref{nIG}. These
conclusions hold even for $\sg>1$. Therefore, our results can be
accurately reproduced by using exclusively $\Vhi$ in \Eref{Vhi2o}.

\renewcommand{\arraystretch}{1.25}
\begin{table}[!t]
\bec\begin{tabular}{|c||c|c|c||c|c||c|c||c|c|c|}\hline
{\sc Model} &  \multicolumn{5}{c||}{\sc Input
Parameters}&\multicolumn{5}{|c|}{\sc Output
Parameters}\\\cline{2-11}
&$n$&$m$&$\na$&$\ck$&$\sgx$&$\ld~(10^{-3})$&$\sgf$&$\ns$&$\as
(10^{-4})$&$r (10^{-3})$\\\hline
Ceccoti-like&$1$&$1$&$2$ &$1$& $82$&$0.0018$ &$1.7$&$0.965$ & $-6.3$& $2.4$\\
Dilatonic&$k/2$&$k$&$1$ &$1$& $61$&$0.0017$ &$2$&$0.966$ & $-6.$& $4.4$\\
No-scale&$3k/2$&$k$&$3$ &$31$& $1$&$3.5$&$0.25$ &$0.965$ & $-6.3$&
$1.6$\\
%&&&&\multicolumn{2}{|c||}{$(m=2)$}&&&\\\cline{5-6}
%
IG Model&$k$&$k$&$2$ &$116$& $1$&$2$&$0.14$& $0.965$ & $-6.3$&
$2.4$\\ With $\mathbb{Z}_k$&&&&&&&&&&\\\hline
\end{tabular}\eec
\hfill \caption[]{\sl\small  Input and output parameters of the
models which are compatible with \Eref{Nhi} for $\Ns=55$ and
\Eref{Prob}. In cases that $n$ and $m$ are not specified
numerically we take $k=2$ for the computation of the parameters
$\ld, \ck,\sgx$ and $\sgf$.} \label{tab2}
\end{table}

Our numerical findings for some representative values of $n, m$
and $\nc$ are presented in \Tref{tab2}. In the first row we
present results associated to a Ceccoti-like model \cite{cec},
which is defined by $\ck=n=m=1$. \Eref{con2} implies that $\na=2$
and not $3$ as in the original model \cite{linde, eno7}. In the
second and third rows we present a dilatonic and a no-scale model
defined by $\na=1$ and $3$, respectively. Therefore, \Eref{con2}
yields a relation between $n$ and $m$. In the last row we show
results concerning the IG model \cite{rena,nIG} with the inflaton
raised to the same exponent $n$ in $W$ and $K_3$ in \eqs{Wn}{K3}.
In this case, \Eref{con2} dictates that $\na=2$. The extended IG
model described in \Sref{fhi} provides the necessary flexibility
to obtain solutions to \Eref{con2}, even with $\na\neq2$, by
selecting appropriately the values of $n$ and $m$, as in the
dilatonic and no-scale cases.

In all cases shown in \Tref{tab2}, our predictions for $\ns, \as$
and $r$ depend exclusively on $\na$, and they are in excellent
agreement with the analytic findings of Eqs.~(\ref{ns}) --
(\ref{resr}). On the other hand, the presented $\ck,\ld, \sgx$ and
$\sgf$ values depend on two of the three parameters $n, m$ and
$\na$. For the values displayed, we take $k=2$. We remark that the
resulting $\ns\simeq0.965$ is close to its observationally central
value; $r$ is of the order $0.001$, and $|\as|$ is negligible.
Although the values of $r$ lie one order of magnitude below the
central value of the present combined \bcp\ and \plk\ results
\cite{gws}, these are perfectly consistent with the $95\%$ c.l.
margin in \Eref{data}. In the first two models, we select $\ck=1$
and so inflation takes place for $\sg\geq1$ whereas for the two
other cases we choose a $\ck$ value so that $\sgx=1$. Therefore,
the presented $\ck$ is the minimal one, in agreement with
\Eref{ckmin}. Finally in all cases, we obtain
$\msn\simeq1.25\cdot10^{-5}$ as anticipated in \Eref{msn}.

The most crucial output of our computation is the stabilization of
$S$ (and $\theta$) during and after inflation. To highlight
further this property, we present in \Fref{fig} the variations of
$\what m^2_{s}/\Hhi^2$ and $\what m^2_{\th}/\Hhi^2$ as functions
of $\sg$ for the inputs shown in the two last rows of \Tref{tab2},
taking $\nb=\na$ and $k=2$. It is evident that $\what
m^2_{s}/\Hhi^2$ and $\what m^2_{\th}/\Hhi^2$ remain larger than
unity for $\sgf\leq\sg\leq\sgx$, where $\sgx$ and $\sgf$ are also
depicted -- the two $\sgx$'s are indistinguishable in
\sFref{fig}{b}. For most $\sg$ values, $\what
m^2_{s}/\Hhi^2\simeq2$ (light gray lines) or $3$ (black lines) for
the no-scale or the IG model with a $\mathbb{Z}_2$ symmetry,
respectively, whereas $\what m^2_{\th}/\Hhi^2\simeq6$ for both
cases. Note, finally, that both $\what m^2_{s}/\Hhi^2$ and $\what
m^2_{\th}/\Hhi^2$ are decreasing functions of $\sg$, and so if
these are larger than unity for $\sg=\sgx$, they remain so for
$\sg<\sgx$ too. This behavior is consistent with the formulae of
Table~1, given that $\fp^2$ in the denominator of $\what
m_{\chi^\al}^2$ decreases with $\sg$.

\section{Conclusions}\label{con}

%%%%%%%%%%%%%%%%%%%%%%%%%%%%%%%%%%%%%%%%%%%%%%%%%%%%%%%%%%%%%%%%%%%%
\begin{figure}[!t]\vspace*{-.12in}
\hspace*{-.19in}
\begin{minipage}{8in}
\epsfig{file=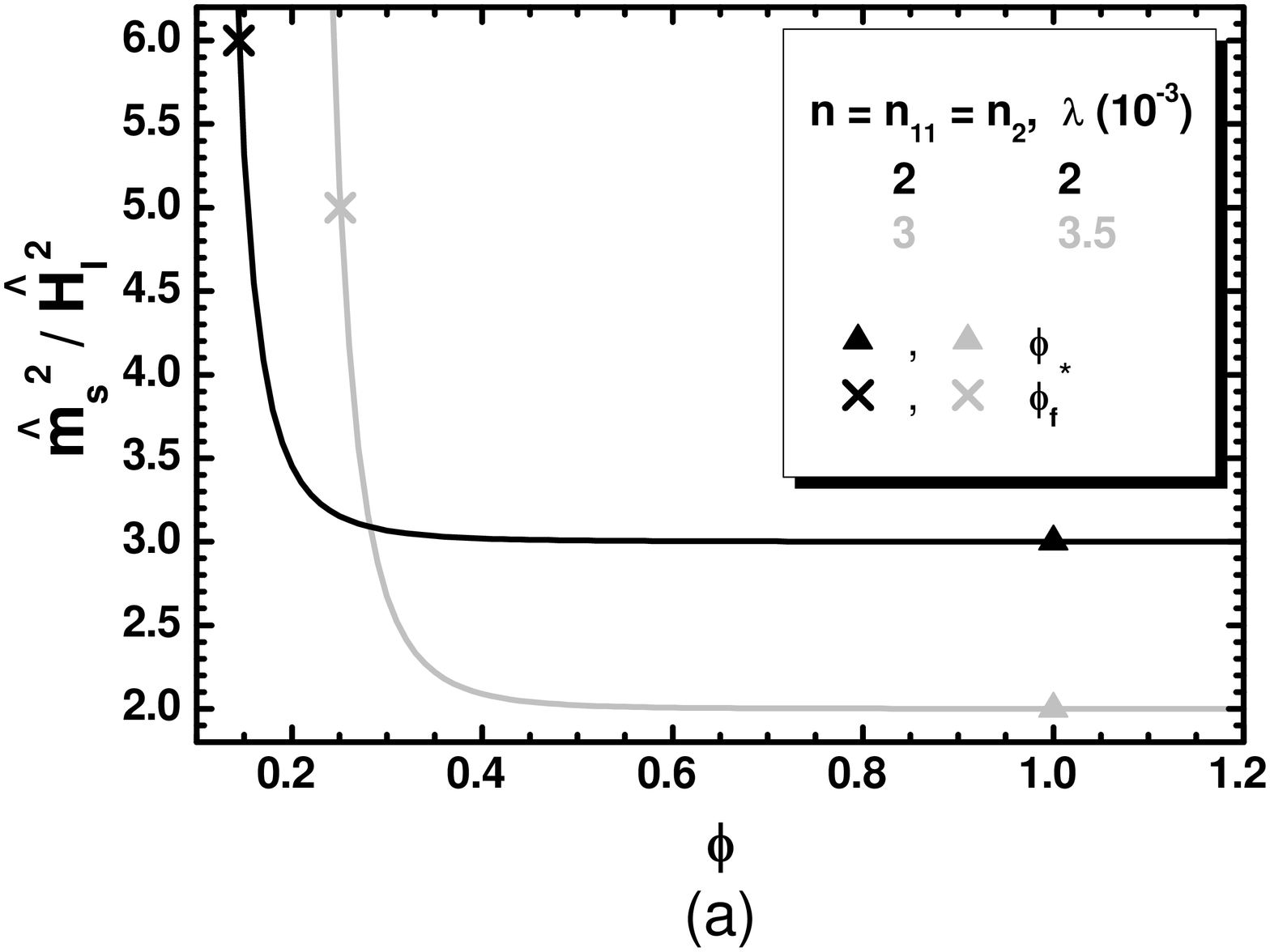,height=3.6in,angle=-90}
\hspace*{-1.2cm}
\epsfig{file=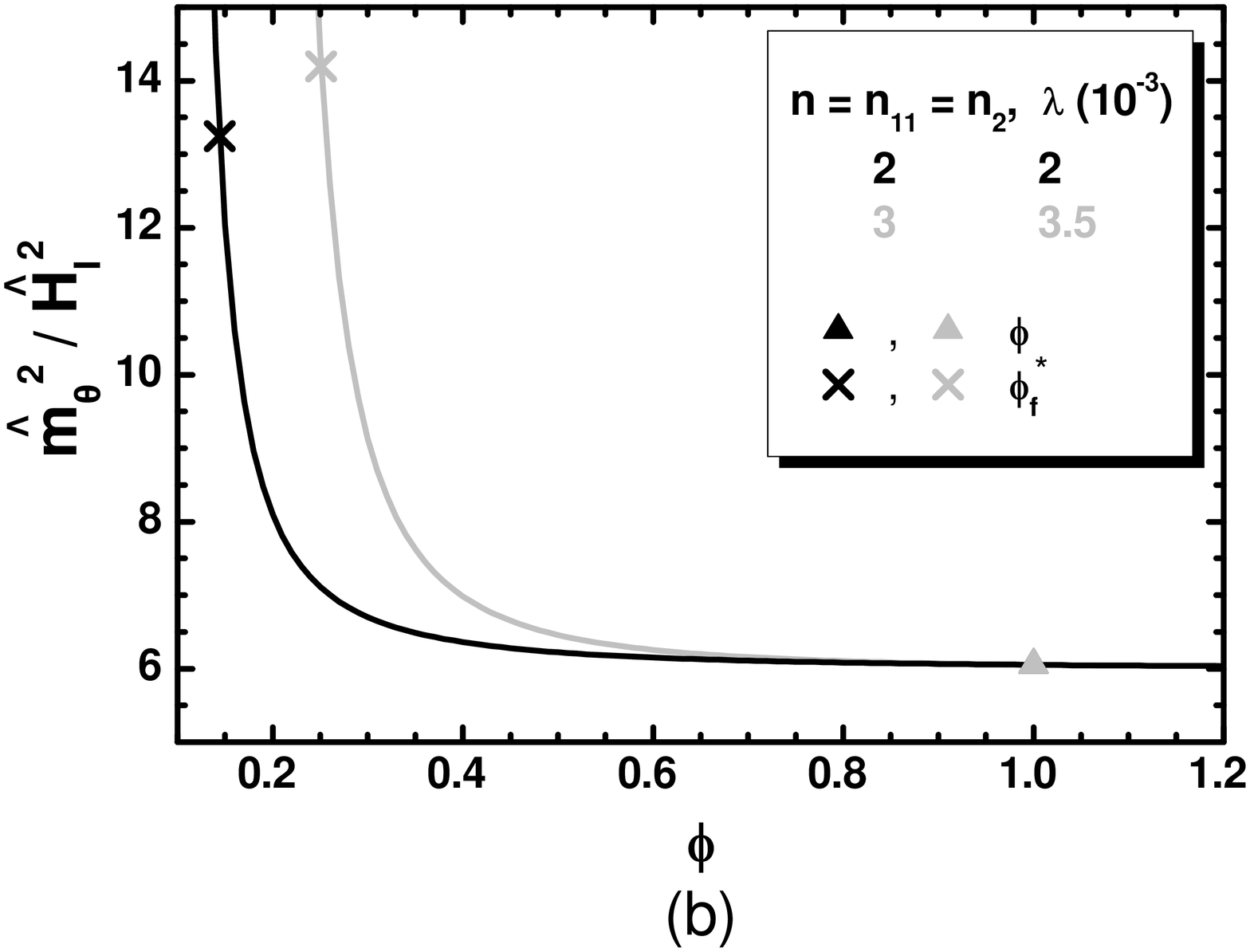,height=3.6in,angle=-90} \hfill
\end{minipage}
\hfill \caption[]{\sl \small The ratios $\what m^2_{s}/\Hhi^2$
{\sffamily\ftn (a)} and $\what m^2_{\theta}/\Hhi^2$ {\sffamily\ftn
(b)} as functions of $\sg$ for $n=m=\na=\nb=2$ and
$\ld=2\cdot10^{-3}$ (black lines) or $m=2$, $n=\nb=\na=3$ and
$\ld=3.5\cdot10^{-3}$ (light gray lines). The values corresponding
to $\sgx$ and $\sgf$ are also depicted.}\label{fig}
\end{figure}
%%%%%%%%%%%%%%%%%%%%%%%%%%%%%%%%%%%%%%%%%

We showed that \str-like inflation can be established in the
context of SUGRA using the superpotential in \Eref{Wn} and the
\Ka\ in \Eref{K3}, which parameterizes the product space
$SU(1,1)/U(1)\times SU(2)/U(1)$. Extending previous work
\cite{rena,nIG}, based on induced gravity, we allow for the
presence of different monomials (with exponents $n$ and $m$) of
the inflaton superfield in $W$ and $K$. Observationally acceptable
inflationary solutions are attained imposing the condition in
\Eref{con2}, which relates the exponents above with the curvature
of the $SU(1,1)/U(1)$ space, $-2/\na$. As a consequence the
inflationary predictions exhibit an attractor behavior depending
exclusively on $\na$. Namely, we obtained $\ns\simeq0.965$ and
$0.001\lesssim r\lesssim0.005$ with negligible $\as$. Moreover,
the mass of the inflaton turns out to be close to
$1.25\cdot10^{-5}$. The accompanying field $S$ is heavy enough and
well stabilized during and after inflation, provided that the
curvature of the $SU(2)/U(1)$ space is such that $0<\nb\leq6$.
Therefore, \str\ inflation realized within this SUGRA setting
preserves its original predictive power. Furthermore it could be
potentially embedded in string theory. If we adopt $\ck\gg1$ and
$n=m>1$, our models can be fixed if we impose two global
symmetries -- a continuous $R$ and a discrete $\mathbb{Z}_n$
symmetry -- in conjunction with the requirement that the original
inflaton takes \sub\ values. The one-loop radiative corrections
remain subdominant and the corresponding effective theories can be
trusted up to $\mP$.

It is argued \cite{2attr} that the models described by \Eref{Vhi1}
for $n=m$ and $\nc=3$ develop one more attractor behavior towards
the $(\ns,r)$'s encountered in the model of quadratic chaotic
inflation. However, this result is achieved only for
transplanckian inflaton values, without preserving the
normalization of $\As$ in \Eref{Prob}. For these reasons we did
not pursue our investigation towards this direction. As a last
remark, we would like to point out that the $S$-stabilization
mechanism proposed in this paper has a much wider applicability.
It can be employed to the models of ordinary \cite{linde1} or
kinetically modified \cite{lazarides, nMkin} non-minimal chaotic
(and Higgs) inflation, without causing any essential alteration to
their predictions. The necessary modifications are to split the
relevant \Ka\ into two parts, replacing the $|S|^2$ depended  part
by the corresponding one included in $K_3$ -- see \Eref{K3} -- and
adjusting conveniently -- as in \Eref{con2} -- the prefactor of
the logarithm including the inflaton in its argument. In those
cases, though, it is not clear if the part of the \Ka\ for the
inflaton sector parameterizes a symmetric \Km\ as in the case
studied here.

%Our present scheme can not be applied only in models introduced in
%\cref{np1} where a term of the form $|S|^2|T|^2$ in $K$ is
%necessary in order to accommodate a sizable increase of the $r$
%values above those obtained here.

\section*{Appendices}

%\newpage
\appendix
\setcounter{equation}{0}
\renewcommand{\theequation}{A.\arabic{equation}}
\renewcommand{\thesubsubsection}{A.\arabic{subsubsection}}

\section{Mathematical Supplement}\label{math}

In this Appendix we review some mathematical properties regarding
the geometrical structure of the $SU(1,1)/U(1)\times SU(2)/U(1)$
\Km. For simplicity we present the case for which $\ck=n=m=1$ in
\eqs{K3}{Wn}. The structure of the $SU(1,1)/U(1)$ coset space
becomes more transparent if we define \cite{lahanas,eno7}
\beq T
=\frac12\frac{1-Z/\sqrt{\na}}{1+Z/\sqrt{\na}}\,\cdot\label{TZ}\eeq
Upon the coordinate transformation above and a \Kaa\
transformation, the model described by the \Ka\
\beq \label{K3t} \wtilde
K_3=-\na\ln\lf1-\frac{|Z|^2}{\na}\rg+\nb\ln\lf1+\frac{|S|^2}{\nb}\rg\eeq
and the superpotential
\beq\label{tWn} \wtilde W=W(1+Z/\sqrt{\na})^{\na}\eeq
is equivalent to the model described by \eqs{Wn}{K3}. The
Riemannian metric associated with $\wtilde K_3$ is given by
\beq\label{ds} ds^2=g_{11}dZdZ^*+g_2dSdS^*\,,\eeq
having a diagonal structure with
\beq\label{gs}
g_{11}=\lf1-|Z|^2/\na\rg^{-2}\>\>\>\mbox{and}\>\>\>g_2=\lf1+|S|^2/\nb\rg^{-2}\,.\eeq
It is straightforward to show that the form of the line element in \Eref{ds}
remains invariant under the transformations
\beq \frac{Z}{\sqrt{\na}}\to
\frac{a_1Z/\sqrt{\na}+b_1}{b_1^*Z/\sqrt{\na}+a_1^*}
\>\>\>\mbox{and}\>\>\> \frac{S}{\sqrt{\nb}}\to
\frac{a_2S/\sqrt{\nb}+b_2}{-b_2^*S/\sqrt{\nb}+a_2^*}\,,
\label{t12}\eeq provided that $|a_1|^2-|b_1|^2=1$ and
$|a_2|^2+|b_2|^2=1$. The \Ka\ $\wtilde K_3$ in \Eref{K3t} remains
invariant under \Eref{t12}, up to a \Kaa\ transformation.

The transformations in \Eref{t12} can be used to define transitive
actions of the $2\times 2$ matrices
\beq
U_1=\mtta{a_1}{b_1}{b_1^*}{a_1^*}\>\>\>\mbox{and}\>\>\>U_2=\mtta{a_2}{b_2}{-b_2^*}{a_2^*}\label{U12}\eeq
on the scalar field manifolds parameterized by $Z$ and $S$ respectively.
These matrices have the properties
\beq U_1^\dag \sigma_3 U_1=\sigma_3\>\>\>\mbox{and}\>\>\>U_2^\dag
U_2=\openone\>\>\>\mbox{with}\>\>\>\sigma_3=\diag\lf1,-1\rg\>\>\>\mbox{and}\>\>\>\openone=\diag\lf1,1\rg\,
,\eeq and so, they provide representations of the $SU(1,1)$ and
$SU(2)$ groups respectively. Now $U_j$ with $j=1,2$ can be written
as $U_j=\wtilde U_j H_j$ (no summation over $j$ is applied), where
the diagonal matrices $H_j=\diag\lf e^{i\th_j}, e^{-i\th_j}\rg$
stabilize the origins of the scalar field manifolds parameterized
by $Z$ and $S$. Thus, the scalar field manifolds are isomorphic to
the coset spaces $SU(1,1)/U(1)$ and $SU(2)/U(1)$. Notice that
\beq \wtilde
U_1=\mtta{\alpha_1}{c_1}{c_1^*}{\alpha_1}\>\>\>\mbox{and}\>\>\>\wtilde
U_2=\mtta{\alpha_2}{c_2}{-c_2^*}{\alpha_2}\label{U12c}\eeq
with $\alpha_j$ real and positive, $\alpha_1^2-|c_1|^2=1$ and
$\alpha_2^2 + |c_2|^2=1$. Therefore, $\wtilde U_j$ with $j=1$  and
$2$ define equivalent parameterizations of the coset spaces
$SU(1,1)/U(1)$ and $SU(2)/U(1)$ respectively.

Finally, applying the formula
\beq \label{Rcrv} R_{K}=g^{-3}\lf g_{,z}g_{,z^*}-gg_{,zz^*}\rg\eeq
for $g=g_{11}$ and $z=Z$ or $g=g_2$ and $z=S$, we find that the
scalar curvatures of the spaces $SU(1,1)/U(1)$ and $SU(2)/U(1)$
are $-2/\na$ and $2/\nb$ respectively.

\setcounter{equation}{0}
\renewcommand{\theequation}{B.\arabic{equation}}
\section{The Effective Cut-off Scale}\label{eff}

A characteristic feature of \FHI compared to conventional
non-minimal chaotic inflation \cite{linde1} is that
perturbative unitarity is retained up to $\mP$, despite the fact
that its implementation with \sub\ $\sg$ values requires
relatively large values of $\ck$ -- see \Eref{ckmin}. To show that
this statement holds in the context of the generalization outlined in
\Sref{fhi}, we extract the ultraviolet cut-off scale $\Qef$ of the
effective theory following the systematic approach of
\cref{riotto}. We focus on the second term in the right-hand side
of \Eref{Saction1} for $\mu=\nu=0$ and $\al=\bt=1$, and we expand
it about $\vev{\phi}$, given by \Eref{dphi}, in terms of $\dphi$.
Our result is written as
\beqs\beq J^2
\dot\phi^2\simeq\lf1-\frac{\sqrt{2\na}}{n}\dphi+\frac{3\na}{2n^2}\dphi^2-
\frac{\sqrt{2}n^{3/2}_{11}}{n^3}\dphi^3+\cdots\rg\dot\dphi^2,\eeq
where we take into account \Eref{con2}. Expanding similarly $\Vhi$
in \Eref{Vhi2o} we obtain
\beq \Vhi\simeq\frac{\na\ld^2\se^2}{2^{\na+1}\ck^{\na}}
\lf1-\sqrt{\frac{\na}{2}}\frac{n+1}{n}\,\dphi+\na\lf1+n\rg\frac{11+7n}{24n^2}\,\dphi^2-\cdots\rg.
\eeq\eeqs
Since the coefficients in the series above are independent of
$\ck$ and of order unity for reasonable $n$ and $\na$ values,
we infer that our models do not face any problem with perturbative
unitarity up to $\mP$.

\def\ijmp#1#2#3{{\sl Int. Jour. Mod. Phys.}
{\bf #1},~#3~(#2)}
\def\plb#1#2#3{{\sl Phys. Lett. B }{\bf #1}, #3 (#2)}
\def\prl#1#2#3{{\sl Phys. Rev. Lett.}
{\bf #1},~#3~(#2)}
\def\rmp#1#2#3{{Rev. Mod. Phys.}
{\bf #1},~#3~(#2)}
\def\prep#1#2#3{{\sl Phys. Rep. }{\bf #1}, #3 (#2)}
\def\prd#1#2#3{{\sl Phys. Rev. D }{\bf #1}, #3 (#2)}
\def\prdn#1#2#3#4{{\sl Phys. Rev. D }{\bf #1}, no. #4, #3 (#2)}
\def\npb#1#2#3{{\sl Nucl. Phys. }{\bf B#1}, #3 (#2)}
\def\npps#1#2#3{{Nucl. Phys. B (Proc. Sup.)}
{\bf #1},~#3~(#2)}
\def\mpl#1#2#3{{Mod. Phys. Lett.}
{\bf #1},~#3~(#2)}
\def\jetp#1#2#3{{JETP Lett. }{\bf #1}, #3 (#2)}
\def\app#1#2#3{{Acta Phys. Polon.}
{\bf #1},~#3~(#2)}
\def\ptp#1#2#3{{Prog. Theor. Phys.}
{\bf #1},~#3~(#2)}
\def\n#1#2#3{{Nature }{\bf #1},~#3~(#2)}
\def\apj#1#2#3{{Astrophys. J.}
{\bf #1},~#3~(#2)}
\def\mnras#1#2#3{{MNRAS }{\bf #1},~#3~(#2)}
\def\grg#1#2#3{{Gen. Rel. Grav.}
{\bf #1},~#3~(#2)}
\def\s#1#2#3{{Science }{\bf #1},~#3~(#2)}
\def\ibid#1#2#3{{\it ibid. }{\bf #1},~#3~(#2)}
\def\cpc#1#2#3{{Comput. Phys. Commun.}
{\bf #1},~#3~(#2)}
\def\astp#1#2#3{{Astropart. Phys.}
{\bf #1},~#3~(#2)}
\def\epjc#1#2#3{{Eur. Phys. J. C}
{\bf #1},~#3~(#2)}
\def\jhep#1#2#3{{\sl J. High Energy Phys.}
{\bf #1}, #3 (#2)}

\end{document}